\documentclass[epj]{svjour}
\usepackage{amssymb}
\usepackage{graphicx}
\usepackage{dcolumn,color}
\usepackage{bm}
\usepackage{epstopdf}
\usepackage{epsfig}
\usepackage{color,framed}
\usepackage{amsmath,amssymb}
\begin{document}
\title{Pure geometric thick $f(R)$-branes: stability and localization of gravity}

\author{Yuan Zhong \and Yu-Xiao Liu}                     
%
\mail{liuyx@lzu.edu.cn}
\institute{Institute of Theoretical Physics, Lanzhou University, Lanzhou 730000, People's Republic of China}
\date{Received: date / Revised version: date}
%
\abstract{We study two exactly solvable five-dimensional thick brane world models in pure metric $f(R)$ gravity. Working in the Einstein frame, we show that these solutions are stable against small linear perturbations, including the tensor, vector, and scalar modes. For both models, the corresponding gravitational zero mode is localized on the brane, which leads to the four-dimensional Newton's law; while the massive modes are nonlocalized and only contribute a small correction to the Newton's law at a large distance.
} 
\maketitle
\section{Introduction}
The idea that our world might be a hyperspace (called brane world) embedded in higher-dimensional space-time (called the bulk) has been intensively considered in the passed two decades~\cite{Akama1982,RubakovShaposhnikov1983,Visser1985,Antoniadis1990,AntoniadisArkani-HamedDimopoulosDvali1998,Arkani-HamedDimopoulosDvali1998a,RandallSundrum1999,RandallSundrum1999a,AntoniadisArvanitakiDimopoulosGiveon2012,YangLiuZhongDuWei2012} (for reviews, see~\cite{Rubakov2001,Csaki2005}). This idea has changed our traditional knowledge toward extra dimensions. In early theories of extra dimensions, namely, the Kaluza-Klein type theories, the extra dimensions are compacted to the Planck scale~\cite{AppelquistChodosFreund1987}. While in brane world scenarios, depend on the model, the radiuses of extra dimensions can be as large as a few TeV$^{-1}$~\cite{Antoniadis1990,AntoniadisArvanitakiDimopoulosGiveon2012,YangLiuZhongDuWei2012}, or several millimeters ~\cite{AntoniadisArkani-HamedDimopoulosDvali1998}, or even be infinitely large~\cite{RubakovShaposhnikov1983,RandallSundrum1999a}.

In one of Randall and Sundrum's brane world scenarios (the RS-II model)~\cite{RandallSundrum1999a}, the authors considered a 3-brane embedded in a five-dimensional anti de Sitter (AdS) space. Due to the nonfactorizable background geometry, the spectrum of four-dimensional gravitons of the RS-II model is consisted by a normalizable zero mode along with a continuum of non-localized massive KK modes. The normalizable zero mode corresponds to the four-dimensional massless graviton and leads to the Newton's law. There is no mass gap between the zero mode and massive modes. So by intuition, the massive modes should cause a large correction to the four-dimensional Newton's law. However, after calculation Randall and Sundrum surprisingly found that the whole continuum of massive modes only contributes a small correction to the Newton's law at a large distance~\cite{RandallSundrum1999a}. In other words, the continuum modes are decoupled.

In the set up of the RS-II model~\cite{RandallSundrum1999a}, the 3-brane has no thickness, and the geometry has a singularity at the location of the brane. To evade this singularity, one can extend the RS-II model by replacing the original 3-brane by a smooth domain wall (called thick brane) generated by a background scalar field~\cite{DeWolfeFreedmanGubserKarch2000,Gremm2000,CsakiErlichHollowoodShirman2000}. Due to the configuration of the domain wall, the bulk is not an AdS$_5$ space now. But the geometry is asymptotically AdS at the infinity of the extra dimension. Thanks to this asymptotic behavior of the geometry, the gravitational zero mode is usually normalizable~\cite{DeWolfeFreedmanGubserKarch2000,Gremm2000,CsakiErlichHollowoodShirman2000}. Besides, the authors of ref.~\cite{CsakiErlichHollowoodShirman2000} found that at least for two mass points localized on the center of the thick brane, the continuum modes are decoupled provided the zero mode is normalizable.

In addition to thick branes generated by scalar fields, there are also thick branes arise from pure geometry. For instance, by replacing the Riemannian geometry into a Weyl-integrable geometry, the authors of refs.~\cite{AriasCardenasQuiros2002,Barbosa-CendejasHerrera-Aguilar2005,Barbosa-CendejasHerrera-Aguilar2006,Barbosa-CendejasHerrera-AguilarReyesSchubert2008} constructed thick branes without introduce an additional matter field. The normalization of the gravitational zero mode as well as the decoupling of the massive Kaluza-Klein (KK) modes are also studied therein.

In this paper, we investigate another alternative for generating thick branes with only geometry. We assume that gravity is not described by general relativity, but by the so-called $f(R)$ theories, where the Lagrangians are proportional to some functions of the scalar curvature $R$. The $f(R)$ theories were created in the study of cosmology~\cite{Buchdahl1970,BarrowOttewill1983,BarrowCotsakis1988}, and are mainly applied in cosmology nowadays (see~\cite{MukhanovKofmanPogosian1987,CognolaElizaldeNojiriOdintsovSebastianiZerbini2008,DeTsujikawa2010,SotiriouFaraoni2010,NojiriOdintsov2011} and references therein). Nevertheless, there are works which devote to embedding branes, either thin~\cite{ParryPichlerDeeg2005,DeruelleSasakiSendouda2008,BalcerzakDabrowski2009,BalcerzakDabrowski2008,HoffDias2011,CaramesGuimaraesHoff2013}, or thick~\cite{AfonsoBazeiaMenezesPetrov2007,LiuZhongZhaoLi2011,BazeiaMenezesPetrovSilva2013,BazeiaLobaoMenezesPetrovSilva2014,XuZhongYuLiu2015,GuGuoYuLiu2014,BazeiaLobaoMenezes2015,BazeiaLobaoLosanoMenezesOlmo2015,YuZhongGuLiu2015}
into various types of $f(R)$ gravities.

Note that, all the thick $f(R)$-branes considered in refs.~\cite{AfonsoBazeiaMenezesPetrov2007,LiuZhongZhaoLi2011,BazeiaMenezesPetrovSilva2013,BazeiaLobaoMenezesPetrovSilva2014,XuZhongYuLiu2015,GuGuoYuLiu2014,BazeiaLobaoMenezes2015,BazeiaLobaoLosanoMenezesOlmo2015,YuZhongGuLiu2015} are generated by a background scalar field. For $f(R)$-branes of this type, the tensor perturbation equation has been derived in ref.~\cite{ZhongLiuYang2011}, but it is still unclear if these models are stable against the vector and especially, the scalar perturbations. To obtain reliable thick $f(R)$-brane models, we must either proof that the solutions found in refs.~\cite{AfonsoBazeiaMenezesPetrov2007,LiuZhongZhaoLi2011,BazeiaMenezesPetrovSilva2013,BazeiaLobaoMenezesPetrovSilva2014,XuZhongYuLiu2015,GuGuoYuLiu2014,BazeiaLobaoMenezes2015,BazeiaLobaoLosanoMenezesOlmo2015,YuZhongGuLiu2015} are also stable against the vector and the scalar perturbations, or to find some new solutions whose stabilities are easier to proof. In this paper we adopt the second path.

According to the well-known Barrow-Cotsakis theorem~\cite{BarrowCotsakis1988}, a pure metric $f(R)$ theory (referred to as the Jordan frame) is conformally equivalent to general relativity minimally coupled with a single canonical scalar field (called the Einstein frame). This equivalence implies the possibility for constructing thick branes without introducing additional matter fields. More importantly, it is much easier to analyze the linear stability of a solution in the Einstein frame.
To the best of our knowledge, however, only refs.~\cite{DzhunushalievFolomeevKleihausKunz2010,LiuLuWang2012} considered thick RS-II brane world solutions in pure $f(R)$ gravity. In~\cite{DzhunushalievFolomeevKleihausKunz2010}, the authors obtained a few numerical solutions. While the first analytical thick brane solution was reported recently in~\cite{LiuLuWang2012}.

In this paper, we derive two analytical thick RS-II brane solutions in pure $f(R)$ theories: one with a triangular $f(R)$ and the other a polynomial $f(R)$. The first solution is equivalent to the one of ref.~\cite{LiuLuWang2012}, despite an apparent difference. The second one is a new solution. These solutions will be presented in the next section. In section \ref{sec3}, we analyze the linear stability of these solutions in the Einstein frame by directly citing the results of refs.~\cite{Giovannini2001a,Giovannini2002}. Then in section \ref{sec4}, we show that the gravitational zero modes correspond to our solutions are normalizable, which implies that the four-dimensional Newton's law can be reproduced on the branes. In section \ref{sec5}, by analyzing the asymptotic behavior of our solutions, we draw a conclusion that for two mass points localized at the vicinity of the brane, the massive KK modes are decoupled, and only lead to small corrections the Newton's law. In the last section, we summarize the main results of this paper.

\section{The model and the solution}
We consider pure $f(R)$ gravity in five-dimensional space-time
\begin{eqnarray}
\label{action}
S=\frac{1}{2\kappa_5^2}\int d^5x\sqrt {-g}f(R),
\end{eqnarray}
where $\kappa_5^2=8\pi G^{(5)}$ is the five-dimensional gravitational coupling constant, and $g=\det (g_{MN})$ is the determinant of the metric. In this paper we only consider the flat and static brane, for which the metric takes the following form:
\begin{eqnarray}
\label{metricy}
ds^2=e^{2A(y)}\eta_{\mu\nu}dx^\mu dx^\nu+dy^2,
\end{eqnarray}
where $e^{2A(y)}$ is the warp factor, $\eta_{\mu\nu}$ is the four-dimensional Minkowski metric, and $y=x^4$ denotes the extra dimension. Throughout this paper,
capital Latin letters $M,N,\cdots=0,1,2,3,5$ and Greek letters $\mu,\nu,\cdots=0,1,2,3$ are used to represent the bulk and brane indices, respectively.

The Einstein equations read
\begin{eqnarray}
\label{EE1}
  f(R)+2f_R\left(4\dot{A}^2+\ddot{A}\right)
  -6\dot{f}_R\dot{A}-2\ddot{f}_R=0,
\end{eqnarray}
and
\begin{eqnarray}
\label{EE2}
  -8f_R\left(\ddot{A}+\dot{A}^2\right)+8\dot{f}_R\dot{A}
  -f(R)=0,
\end{eqnarray} where $f_R\equiv df(R)/dR$, and the over dots denote the derivatives with respect to $y$.
By eliminating $f(R)$, one immediately obtains the following equation:
\begin{eqnarray}
\label{EEPure}
\ddot{f}_R-\dot{f}_R \dot{A} +3f_R\ddot{A}=0.
\end{eqnarray}

For a specified $A(y)$, eq.~(\ref{EEPure}) is a second-order differential equation for $f_R(y)$. In case $A(y)$ takes a simple mathematical form, it is possible to solve $f_R(y)$ analytically. By inserting $A$ and $f_R$ back into eq.~(\ref{EE2}), one can easily get the solution of $f(R)$ as a function of $y$. Note that for the metric (\ref{metricy}), the scalar curvature $R$ is related to $y$ via the following equation
\begin{eqnarray}
\label{scalarR}
R=-20 \dot{A}^2-8 \ddot{A}.
\end{eqnarray}
Once we get the expression of $R(y)$, it is not difficult to rewrite $f_R$ and $f(R)$ as functions of $R$.

Instead of starting with a simple $f(R)$, we prefer to begin with a simple $A(y)$. For instance, we consider
\begin{eqnarray}
A&=&-n \ln (\cosh (k y)),
\end{eqnarray}
with $n$ a dimensionless positive constant, and $k$ another positive constant with the dimension of length inverse. It is convenient for us to introduce a dimensionless variable $w=ky$. In terms of $w$, the scalar curvature takes a simple form:
\begin{eqnarray}
R=8 n k^2 \textrm{sech}^2(w)-20 n^2 k^2 \tanh^2(w),
\end{eqnarray}
from which we can express $w$ in terms of $R$ for an arbitrary $n$:
\begin{eqnarray}
\label{omega}
w(R)=\pm\textrm{arcsech}\left[\frac{\sqrt{20 n^2+R/k^2}}{\sqrt{8n+20 n^2}}\right].
\end{eqnarray}
The above equation makes it possible for us to get the analytical expression of $f(R)$, at least for some special values of $n$.
\subsection{Case 1: $n=1$, triangular $f(R)$}
We first consider the simplest case with $n=1$, and
\begin{eqnarray}
\label{warp}
A&=&-\ln (\cosh (w)).
\end{eqnarray}

Substituting eq.~(\ref{warp}) into eq.~(\ref{EEPure}), and only keep the symmetric solution, one immediately obtains
\begin{equation}
\label{solutfR}
f_R=\cosh \left(\alpha(w)\right),
\end{equation}
where the function $\alpha (w)$ is defined as
\begin{eqnarray}
\alpha (w)\equiv 2 \sqrt{3} \arctan \left(\tanh \left(\frac{w}{2} \right)\right).
\end{eqnarray}
From eq.~(\ref{EE2}), one can easily obtain the solution of $f(R(w))$:
\begin{eqnarray}
\label{10}
f(R(w))&=&4 k^2 (3-\cosh (2 w))\textrm{sech}^2(w) \cosh \left(\alpha (w)\right)\nonumber\\
&-&8 \sqrt{3} k^2 \textrm{sech}(w)\tanh (w) \sinh \left(\alpha (w)\right).
\end{eqnarray}
Using eq.~(\ref{omega}), we get
\begin{eqnarray}
\label{SolutionF}
f(R)&=&\frac{4}{7} \left(6 k^2+R\right)\cosh (\alpha(w(R)) )
\nonumber\\
&-&\frac{2}{7} k^2 \sqrt{480-\frac{36 R}{k^2}-\frac{3 R^2}{k^4}} \sinh (\alpha(w(R)) ).
\end{eqnarray}
Note that $f(R(w))$ is an even function of $w$, so it makes no difference in choosing between the plus sign solution or the other in eq.~(\ref{omega}). Therefore, eqs.~(\ref{warp}) and (\ref{SolutionF}) constitute the first analytically solvable $f(R)$ brane model.

Note that in ref.~\cite{LiuLuWang2012}, the authors also investigated thick RS-II brane solution in pure metric $f(R)$ gravity with the same warp factor (\ref{warp}). They also obtained an analytical expression of $f(R)$. Despite the difference in the mathematical expressions, it can be shown that both solutions are equivalent. The authors of~\cite{LiuLuWang2012} have shown that this solution is stable under tensor perturbation. In the next section, we will prove that this solution is also stable under scalar and vector perturbations.


\subsection{Case 2: $n=20$, polynomial $f(R)$}
The second analytically solvable model appears when $n=20$:
\begin{eqnarray}
\label{warp2}
A&=&-20\ln (\cosh (w)).
\end{eqnarray}
In this case, the symmetric solution of $f_R$ takes the form
\begin{equation}
\label{solutfR2}
f_R=1+30 \tanh^2(w)+65 \tanh^4(w).
\end{equation}
Using the same procedure as  the last subsection, we obtain a simple polynomial solution
\begin{equation}
\label{solutf2}
f(R)=\Lambda+c_1 R-\frac{c_2 }{ k^2}R^2+\frac{c_3 }{ k^4}R^3,
\end{equation}
where
$\Lambda=-\frac{377600 k^2}{7803}$ is the cosmological constant, while $c_1=\frac{4196 }{2601}$,
$c_2=-\frac{83 }{41616 }$,
$c_3=\frac{13 }{39951360 }$ are dimensionless constants.

\section{Linear perturbations and stability of the solutions}
\label{sec3}
In this section, we consider small metric perturbations around the solutions we obtained in the previous section. Our aim is to show that both of the solutions are stable against the metric perturbations to the linear order.

It is well-known that a pure $f(R)$ gravity is conformally equivalent to a theory with a minimally coupled scalar in Einstein's gravity~\cite{BarrowCotsakis1988}. The linear perturbations of the later case has been extensively investigated in literature~\cite{Giovannini2001a,Giovannini2003,Giovannini2002}. Thus, it is more convenient to discuss the stability of our solution in the Einstein frame.
\subsection{The Einstein frame}
Firstly, we define a new variable $z$, such that $dz=e^{-A}dy$. In terms of $z$, the metric can be written as
\begin{eqnarray}
g_{MN}=e^{2A(z)}\eta_{MN}.
\end{eqnarray}

Then, we introduce a conformal transformation
\begin{equation}
\tilde{g}_{MN}=\Omega(z)^2{g}_{MN},
\end{equation}
where $\Omega(z)$ is a function of $z$.
From now on, we will always use a tilde to denote a quantity in the Einstein frame.
Obviously, $\tilde{g}_{MN}$ is conformally flat:
\begin{equation}
\label{metricar}
\tilde{g}_{MN}=\tilde{a}(z)^2{\eta}_{MN}.
\end{equation}
Here $\tilde{a}(z)\equiv e^{A(z)}\Omega$, which will used in next subsection.

Under the conformal transformation, the Ricci scalar transforms as~\cite{Carroll2004}
\begin{equation}
\label{conformalR}
R=\Omega^2\tilde{R}+8\tilde{g}^{MN}\Omega(\tilde{\nabla}_M\tilde{\nabla}_N\Omega)
-20\tilde{X},
\end{equation}
where $\tilde{\nabla}_M$ is the covariant derivative defined by the conformal metric $\tilde{g}_{MN}$, and $\tilde{X}\equiv \tilde{g}^{MN}\tilde{\nabla}_M\Omega\tilde{\nabla}_N\Omega$. To continue, let us first rewrite the original gravitational action (\ref{action}) as
\begin{eqnarray}
\label{conformalS}
 S = \int {d^5}x\sqrt { - \tilde g} \left(\frac{{{}{f_R}}}{{2\kappa _5^2}}\Omega ^{ - 5}R - {\Omega ^{ - 5}}\sigma\right),
\end{eqnarray}
where
\begin{equation}
\sigma\equiv \frac{{{f_R}R - f(R)}}{{2\kappa _5^2}}.
\end{equation}
At this step, we only used the relation $\sqrt{-g}=\Omega^{-5}\sqrt{-\tilde g}$.
Next, we substitute eq.~(\ref{conformalR}) into eq.~(\ref{conformalS}), and take ${f_R} = {\Omega ^3}$, such that
\begin{eqnarray}
&&  S= \int {d^5}x\sqrt { - \tilde g} \left\{ \frac{1}{{2\kappa _5^2}}[\tilde R - 12{\Omega ^{ - 2}}\tilde{X} ] - \frac\sigma{\Omega ^{ 5}}\right\}.
\end{eqnarray}
By defining
\begin{equation}
\phi  =2 \sqrt {\frac{{3}}{{\kappa _5^2}}} \ln \Omega,\quad
V(\phi)={\Omega^{- 5}}\sigma,
\end{equation}
one can finally simplify the action as
\begin{equation}
\label{actionSG}
S = \int {d^5}x\sqrt { - \tilde g} \left\{ \frac{1}{{2\kappa _5^2}}\tilde R - \frac{1}{2}\tilde{X}  -V(\phi)\right \}.
\end{equation} This action describes a minimally coupled scalar field in Einstein's gravity. The linearization of thick brane system with action (\ref{actionSG}) and metric (\ref{metricar}) has been thoroughly studied in~\cite{Giovannini2001a,Giovannini2003}, where the metric perturbations are classified into tensor, vector, and scalar modes. Each type of these modes evolves independently, and none of the perturbation equations relies on the explicit form of $V(\phi)$.
\subsection{Quadratical actions and stability}
Now we consider the linearization of a system defined by the action (\ref{actionSG}) along with the metric (\ref{metricar}). We need to consider perturbations comes from both the scalar field $\phi$ and the metric $\tilde{g}_{MN}$, denoted by $\delta\phi$ and $\delta \tilde{g}_{MN}\equiv \tilde{a}^2(z) h_{MN}$, respectively. To obtain the equations for linear perturbations, one can expand the action (\ref{actionSG}) to the second order of $\delta\phi$ and $h_{MN}$. The result can be found in refs.~\cite{Giovannini2001a,Giovannini2003,Giovannini2002}, but here we use the one of ref.~\cite{ZhongLiu2013}:
\begin{eqnarray}
&& {{S}^{(2)}}  =
  \frac{1}{2}\int {d^5}x{\tilde{a}^{3}}\bigg\{
   {\partial _M}{h_{NP}}{\partial ^P}{h^{MN}} - {\partial ^M}h{\partial ^N}{h_{MN}}\nonumber \\
   &+& 2\kappa _5^2\bigg[{\tilde{a}^2}\frac{\partial^2 V}{\partial\phi^2}{(\delta \phi )^2} + 2{h^{Mz}}\phi '{\partial _M}\delta \phi  + \phi 'h'\delta \phi   \nonumber\\
   &-& {\partial ^M}\delta \phi {\partial _M}\delta \phi \bigg]-\frac{1}{2}{\partial _P}{h_{MN}}{\partial ^P}{h^{MN}} + \frac{1}{2}{\partial ^P}h{\partial _P}h\nonumber \\
   &+& 3\frac{{\tilde{a}'}}{\tilde{a}}(h{\partial ^\mu }{h_{\mu z}} - {h_{zz}}h')\bigg\},
\end{eqnarray}
where
$\partial^M=\eta^{MN}\partial_N$, $\partial^\mu=\eta^{\mu\nu}\partial_\nu$, $h=\eta^{MN}h_{MN}$, and the primes represent the derivatives with respect to $z$. Note that in this subsection, all the upper indices $\mu$, $\nu$ (or $M,N$) are raised by the Minkowski metric $\eta^{\mu\nu}$ (or $\eta^{MN}$). {Note that for pure gravity around Minkowski background ($\tilde{a}=1$), $S^{(2)}$ reduces to the well-known Fierz-Pauli action~\cite{FierzPauli1939}.}

Following the procedures in ref.~\cite{ZhongLiu2013}, we introduce the scalar-tensor-vector (STV) decompositions for the metric perturbation\footnote{The STV decomposition method was firstly introduced in cosmology by Bardeen~\cite{Bardeen1980}, and now is a widly accepted method in dealing with cosmological perturbations~\cite{KodamaSasaki1984,MukhanovFeldmanBrandenberger1992,Weinberg2008}. This method can also be extended in the study of brane world perturbations~\cite{Giovannini2001a,Giovannini2003,Giovannini2002,ZhongLiu2013}.}:
\begin{subequations}
\label{decomposition}
\begin{eqnarray}
{h_{\mu z}} &=& {\partial _\mu }F + {G_\mu },\\
{h_{\mu \nu }} &=& {\eta _{\mu \nu }}\varphi + {\partial _\mu }{\partial _\nu }B + {\partial _{\mu }}{C_{\nu }}+{\partial _{\nu }}{C_{\mu }} + {D_{\mu \nu }},
\end{eqnarray}
\end{subequations}
where $C_\mu, G_\mu$ are transverse vector perturbations:
\begin{eqnarray}
\partial^\mu C_\mu=0=\partial^\mu G_\mu,
\end{eqnarray}
and $D_{\mu \nu }$ denotes the tensor perturbation, which is transverse and traceless (TT):
\begin{eqnarray}
\partial^\nu D_{\mu \nu }=0=D^\mu_\mu.
\end{eqnarray}

The STV decomposition {enables one to decompose ${S}^{(2)}$ into three independent parts:
\begin{eqnarray}
{S}^{(2)}=S_{\textrm{v}}^{(2)}+S_{\textrm{t}}^{(2)}+S_{\textrm{s}}^{(2)}.
\end{eqnarray}
Each type of perturbation evolves independently, and therefore, can be analyzed separately.
}
The vector and tensor sections are
\begin{eqnarray}
\label{37}
  S_{\textrm{v}}^{(2)}& =&\frac{1}{2}\int {d^5}x {\hat{v}^\mu }{\square ^{(4)}}{\hat{v}_\mu },   \\
\label{eq38}
  S_{\textrm{t}}^{(2)}
  &= &\int {d^5}x\frac{\hat{D}^{\mu \nu }}{4}
  \bigg\{\square ^{(4)}{\hat{D}_{\mu \nu }} + \hat{D}_{\mu \nu }'' - \frac{{({\tilde{a}^{\frac{3}{2}}})''}}{{{\tilde{a}^{\frac{3}{2}}}}}{\hat{D}^{\mu \nu }}\bigg\},
\end{eqnarray}
correspondingly, where
$\square^{(4)}=\eta^{\mu\nu}\partial_\mu\partial_\nu$.
The normal modes of the vector and the tensor perturbations are
\begin{equation}
\hat{v}^\mu={\tilde{a}^{\frac{3}{2}}}({G_\mu } - C_\mu'),\quad {\hat{D}^{\mu \nu }} = {\tilde{a}^{\frac{3}{2}}}{D^{\mu \nu }},
\end{equation}
respectively.

{The second-order action of scalar perturbations is more involved, it is composed by two parts \cite{ZhongLiu2013}: $S_{\textrm{s}}^{(2)}={S_{\textrm{s-1}}^{(2)}}+{S_{\textrm{s-2}}^{(2)}}$, where}
\begin{eqnarray}
\label{Sconstraint}
{{S_{\textrm{s-1}}^{(2)}}}=\int {d^5}x {\tilde{a}^3}
 \Big\{ 3\frac{{\tilde{a}'}}{\tilde{a}}{h_{zz}} - 3\varphi' - 2\kappa _5^2\phi '\delta \phi \Big\} {\square ^{(4)}}\psi,
\end{eqnarray} with $\psi\equiv F-\frac12B'$, and
\begin{eqnarray}
\label{SDynamical}
 &&{{S_{\textrm{s-2}}^{(2)}}}
 =\frac{1}{2}\int {d^5}x{\tilde{a}^{ 3}} \Big\{
 - 3\varphi{\square ^{(4)}}\varphi
 - 3{h_{zz}}{\square ^{(4)}}\varphi
  + 6\varphi'\varphi' \nonumber \\
   &-& 3\frac{{\tilde{a}'}}{\tilde{a}}{h_{zz}}(h_{zz}' + 4\varphi')
  +2\kappa _5^2\Big[\delta \phi \square ^{(4)}\delta \phi
   + {\tilde{a}^2}{\frac{\partial^2 V}{\partial\phi^2}}{(\delta \phi )^2} \nonumber \\
   &+& 2{h_{zz}}\phi '\delta \phi'
    + \phi '(h_{zz}' + 4\varphi')\delta \phi
   - (\delta \phi' )^2\Big] \Big\}.
\end{eqnarray}
{The variation ${\delta {S_{\textrm{s-1}}^{(2)}}}/{\delta \psi}=0$ leads to }the following constraint equation
\begin{eqnarray}
\label{hrr}
3\frac{{\tilde{a}'}}{\tilde{a}}{h_{zz}} - 3\varphi' - 2\kappa _5^2\phi '\delta \phi=0.
\end{eqnarray}
Using this equation, one can eliminate $h_{zz}$ in the action (\ref{SDynamical}). After a simplification, one finally obtains~\cite{ZhongLiu2013}
\begin{equation}
\label{eqscalar}
{S_{\textrm{s-2}}^{(2)}} = \int d^5x\hat{\mathcal{G}}
 \left\{
   {\square ^{(4)}}\hat{\mathcal{G}}+ \hat{\mathcal{G}}'' -\frac{\theta''}{\theta}\hat{\mathcal{G}}
\right\}.
\end{equation}
Here, $\hat{\mathcal{G}}$ is a gauge invariant variable defined by
\begin{equation}
\hat{\mathcal{G}}=\frac{\kappa_5}{2}\tilde{a}^{3/2}
 \Big(2\delta \phi  - \frac{{\phi '\tilde{a}}}{\tilde{a}'}\varphi \Big),
\end{equation}
and $\theta$ is a function defined as
\begin{equation}
\label{z}
\theta=\tilde{a}^{3/2}\frac{\phi '\tilde{a}}{\tilde{a}'}.
\end{equation}

{From quadratic actions \eqref{37}, \eqref{eq38} and \eqref{eqscalar}, one can easily obtain the linear perturbation equations via the Hamiltonian variation principle
\begin{eqnarray}
\delta S_{\textrm{v}}^{(2)}/\delta {\hat{v}_\mu } &=& 0,\\
\delta S_{\textrm{t}}^{(2)}/\delta {\hat{D}_{\mu\nu} } &=& 0, \\
\delta S_{\textrm{s-2}}^{(2)}/\delta {\hat{\mathcal{G}}} &=& 0,
\end{eqnarray}
and the final results are~\cite{ZhongLiu2013} (see also~\cite{Giovannini2001a}):}
\begin{eqnarray}
&&\textrm{vector:}\quad \square ^{(4)}{\hat{v}_\mu }=0,\\
\label{eqtensor}
&&\textrm{tensor:}\quad \square ^{(4)}{\hat{D}_{\mu \nu }} + \hat{D}_{\mu \nu }'' - \frac{{({\tilde{a}^{\frac{3}{2}}})''}}{{{\tilde{a}^{\frac{3}{2}}}}}{\hat{D}_{\mu \nu }}=0,\\
&&\textrm{scalar:}\quad {\square ^{(4)}}\hat{\mathcal{G}}+ \hat{\mathcal{G}}'' -\frac{\theta''}{\theta}\hat{\mathcal{G}}=0.
\end{eqnarray}
{Note that the tensor perturbation equation \eqref{eqtensor} has also been derived directly without using conformal transformation by the present authors~\cite{ZhongLiuYang2011}.} Obviously, the normal mode of the vector perturbations has only zero mode. Therefore, our solutions are stable against the vector perturbations. For the tensor and scalar modes, we introduce the following decompositions
\begin{eqnarray}
&&\hat{D}_{\mu\nu}(x^\lambda,z)=\epsilon_{\mu\nu}e^{ip_\lambda x^\lambda} \rho_p(z),\nonumber\\
&&\hat{\mathcal{G}}(x^\lambda,z)=e^{iq_\lambda x^\lambda} \Phi_q(z),
\end{eqnarray} where $\epsilon_{\mu\nu}$ is the TT polarization tensor.

It is not difficult to show that $\rho_p(z)$ and $\Phi_q(z)$ satisfy the following equations
\begin{eqnarray}
\label{hatD}
&&\mathcal{A}_t\mathcal{A}_t^\dagger\rho_p=m_p^2\rho_p,\\
\label{hatG}
&&\mathcal{A}_s\mathcal{A}_s^\dagger \Phi_q=M_q^2\Phi_q,
\end{eqnarray} where $ m_p^2=-p^\mu p_\mu$, $ M_q^2=-q^\mu q_\mu$, and
\begin{eqnarray}
\label{Adagger}
\mathcal{A}_t&=&\frac{d}{dz}+\frac{({\tilde{a}^{\frac{3}{2}}})'}{\tilde{a}^{\frac{3}{2}}},\\
\mathcal{A}_s&=&\frac{d}{dz}+\frac{\theta'}{\theta}.
\end{eqnarray}
In the theory of supersymmetric quantum mechanics, the common structure of eqs.~(\ref{hatD}) and (\ref{hatG}) ensures that both $m_p^2$ and $M_q^2$ are semi-positive definite, namely, $m_p^2, M_q^2\geq 0$ for all $p$ and $q$. Therefore, our solutions are also stable against the tensor and scalar perturbations.

Note that in order to make the variables $\tilde{a}$, $\phi$ and $z$ well defined, our solutions must satisfy $f_R(y)>0$ for $y\in (-\infty,+\infty)$. One can easily show that our solutions (\ref{solutfR}) and (\ref{solutfR2}) satisfy this requirement.

\section{The normalization of the tensor zero mode}
\label{sec4}

Equation~(\ref{hatD}) is in fact a Shr\"odinger-like equation
\begin{eqnarray}
\label{schrotensor}
-\rho_p'' + W(z)\rho_p=m_p^2\rho_p,
\end{eqnarray}
where the effective potential $W(z)$ reads
\begin{eqnarray}
W(z)&=&\frac{{({\tilde{a}^{\frac{3}{2}}})''}}{{{\tilde{a}^{\frac{3}{2}}}}}
=\frac{{({e^{\frac{3}{2}A}}f_R^{\frac{1}{2}})''}}{{e^{\frac{3}{2}A}}f_R^{\frac{1}{2}}}\nonumber\\
&=&\frac34\frac{a'^2}{a^2}
      +\frac32\frac{a''}{a}
      +\frac32\frac{a' f_R'}{a f_R}
      -\frac14\frac{f_R'^2}{f_R^2}
      +\frac12\frac{f_R''}{f_R}.
\end{eqnarray}
This expression consists with the result derived in ref.~\cite{ZhongLiuYang2011}.

The spectrum of the tensor KK modes $\rho_p$ determines the effective four-dimensional gravity. Let us start with the zero mode $\rho_0$ with $m_0=0$. A normalizable $\rho_0$ leads to the four-dimensional Newton's Law~\cite{RandallSundrum1999a,Giovannini2001a}. Besides, the four-dimensional Planck constant is finite only when $\rho_0$ is normalizable~\cite{Giovannini2001a}.

From eq.~(\ref{hatD}) we know that the zero mode of the tensor perturbation takes the form
\begin{eqnarray}
\rho_{0}(z)=\mathcal{N}_T \tilde{a}^{3/2}
 = \mathcal{N}_T {{e^{\frac{3}{2}A}}f_R^{\frac{1}{2}}},
\end{eqnarray}
where $\mathcal{N}_T$ is the normalization constant.
The tensor zero mode $\rho_{0}$ is normalizable provided
\begin{eqnarray}
1&=&\int_{-\infty}^{+\infty} dz|\rho_{0}(z)|^2=\mathcal{N}_T^2 \int_{-\infty}^{+\infty} dz f_R(z) e^{3A(z)}\nonumber\\
&=&\frac{\mathcal{N}_T^2}k\int_{-\infty}^{+\infty} dw f_R(w) e^{2A(w)}.
\end{eqnarray}
Here we have used the relation $dz=e^{-A}\frac{dw}k$. For both of our solutions, the above integration can be done analytically.

For the triangular model (\ref{SolutionF}), the integration gives
\begin{equation}
\frac{\mathcal{N}_T^2}{2k} \cosh\left(\frac{\sqrt{3} \pi }{2}\right)=1,
\end{equation} or
\begin{equation}
\mathcal{N}_T \approx 1.953\sqrt{k}.
\end{equation}
Similarly, for the polynomial model (\ref{solutf2}) we get
\begin{eqnarray}
\mathcal{N}_T\approx 0.857\sqrt{k}.
\end{eqnarray}
Thus, for both of our solutions, the gravitational zero mode is normalizable and can be localized on the brane, which results in the familiar Newton's law on the brane.

\section{Correction to the Newton's law}
\label{sec5}

To obtain an acceptable four-dimensional gravity, we have to require the massive modes $\rho_p$ with $p>0$ do not lead to unacceptably large corrections to the four-dimensional Newton's law (In this case, we also say that the massive modes are decoupled). For simplicity, we follow the study of ref.~\cite{CsakiErlichHollowoodShirman2000} and only consider two massive points $\mu_1$ and $\mu_2$ located at $z=0$. We denote the distance between $\mu_1$ and $\mu_2$ as $r$.

As have been addressed in ref.~\cite{CsakiErlichHollowoodShirman2000}, the asymptotic behavior of the effective potential $W(z)$ at $|z|\to\infty$  determines not only the localization of the zero mode, but also the decoupling of the massive modes. The localization of the zero mode requires that $W(|z|\to\infty)\geq 0$. If $W(|z|\to\infty)> 0$, namely, there is a gap between the zero mode and the excited states, then  we will obtain exponentially suppressed corrections to the Newton's law. The most interesting case is $W(|z|\to\infty) \to 0$. In this case, the scattering states start at $m=0$, and the decoupling of the massive modes becomes a delicate issue. An important result of ref.~\cite{CsakiErlichHollowoodShirman2000} states that if the potential $W(z) \to \beta(\beta + 1)/z^2$ as $|z|\to \infty$, the massive modes will contribute a correction $\Delta U\propto 1/r^{2\beta}$ to the Newton's law at large distance (see also~\cite{BazeiaGomesLosano2009}).

\begin{figure}
\begin{center}
\includegraphics[width=0.45\textwidth]{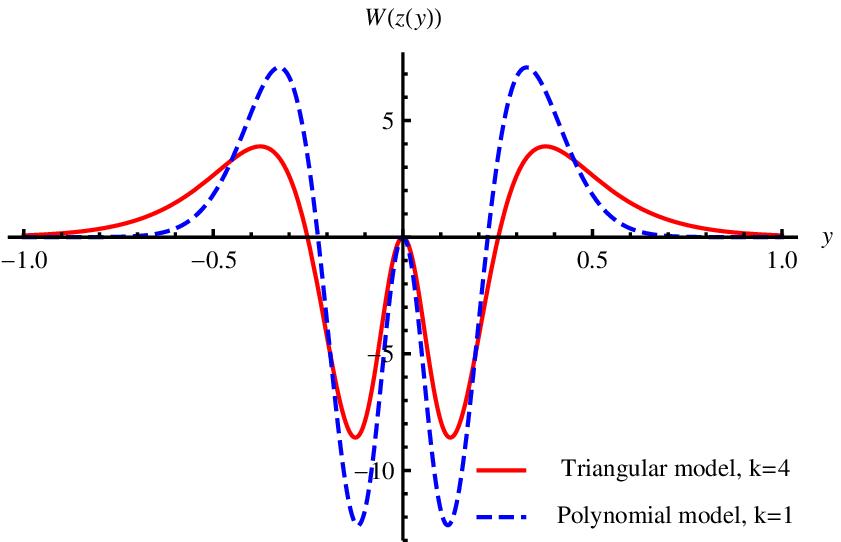}
\end{center}
\caption{Plots of the effective potential $W(z(y))$.} \label{figureWy}
\end{figure}

As depicted in figure~\ref{figureWy}, the effective potentials corresponding to both of our solutions approach to zero as $|y|\to\infty$. They have the same asymptotic behavior in the $z$ coordinate too, as the coordinate transformation is simply a redefinition of $y$, and the shape of $W$ will not change. Thus, our residual task is to prove that $z^2W(z)=\beta(\beta + 1)$ is a constant as $|z|\gg 1$, and to find out the exact values of $\beta$ corresponding to our solutions.

In fact, if $z^2W(z)$ approaches to a constant in the $z$ coordinate, it should have the same asymptotic behavior in the $w$ coordinate, in which all the quantities have analytical forms. For instance, in the $w$ coordinate, the effective potential is expressed as
\begin{eqnarray}
W(z(w))=k^2\frac{{e^{A(w)}\partial_w\left(e^{A(w)}\partial_w({e^{\frac{3}{2}A}}
f_R^{\frac{1}{2}})\right)}}{{e^{\frac{3}{2}A}}f_R^{\frac{1}{2}}},
\end{eqnarray}
and the variable $z$ reads
\begin{eqnarray}
z(w)=\int_0^w d\bar{w}\frac{e^{-A(\bar{w})}}k.
\end{eqnarray}

\begin{figure}
\begin{center}
\includegraphics[width=0.45\textwidth]{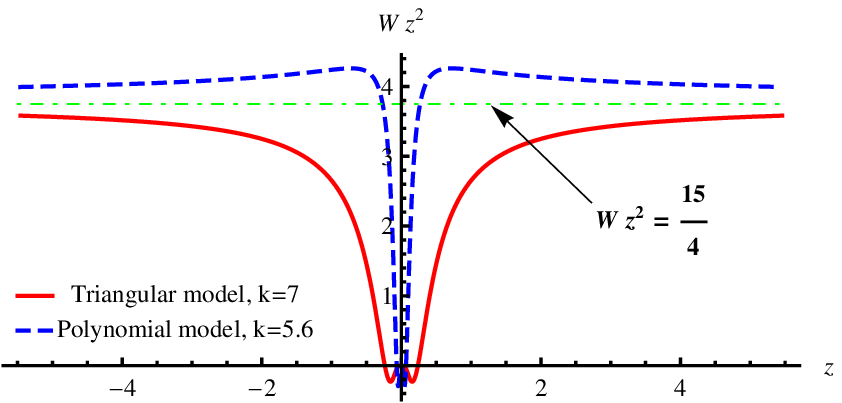}
\end{center}
\caption{Plots of $z^2W(z)$ as a function of $z$. As $|z|\to\infty$, $z^2W(z)\to \frac{15}4$.} \label{figureWz2}
\end{figure}

For both of our solutions, $z^2W(z)$ can be analytically obtained. Instead of writing down the explicit expressions, we show in figure~\ref{figureWz2} that for both of our solutions\\ $\lim_{|z|\to\infty}z^2W(z)=\frac{15}4$, namely $\beta=3/2$. Thus, the corrections to the Newtonian potential $\Delta U\propto 1/r^{3}$ are suppressed at large $r$ for both of our models.

\section{Conclusions}

In this paper, we studied two analytically solvable thick RS-II brane world models in pure metric $f(R)$ gravity theories. Instead of starting with simple forms of $f(R)$, we began with two simple types of metric solutions, and derived the analytical forms of $f(R)$. We obtained two types of $f(R)$ gravities: a triangular one and a polynomial one.

Then we studied the stability of our solutions against small linear perturbations, including tensor, vector, and scalar perturbations. We found that the solutions are stable against all types of perturbations.

In the end, we considered the reproduction of the four-dimensional Newtonian gravity. We first demonstrated that the tensor zero modes are normalizable and localized on the brane for both of our models; so the well-known Newton's gravitational law can be reproduced. Then we showed that for two static mass points localized at $z=0$, all the massive tensor modes contribute a suppressed correction $\Delta U\propto 1/r^{3}$ to the Newton's law. Therefore, the massive tensor modes are decoupled.

This work compensates the studies of ref.~\cite{LiuLuWang2012} by adding a new analytical solution and offering a complete discussion on the stability of the solutions. Our procedures for finding analytical solutions might also be useful for cosmologists.\\

\section*{Acknowledgments}

This work was supported by the National Natural Science Foundation of China (Grant No. 11375075 and No. 11522541), and the Fundamental Research Funds for the Central Universities (Grant No. lzujbky-2015-jl1). Y.Z. was also supported by the scholarship granted by the Chinese Scholarship Council (CSC).


\begin{thebibliography}{10}
\providecommand{\url}[1]{\texttt{#1}}
\providecommand{\urlprefix}{URL }
\providecommand{\eprint}[2][]{\url{#2}}

\bibitem{Akama1982}
K.~Akama, Lect.Notes Phys. \textbf{176}, 267 (1982), \eprint{hep-th/0001113}

\bibitem{RubakovShaposhnikov1983}
V.~A. Rubakov, M.~E. Shaposhnikov, Phys. Lett. B \textbf{125}, 136 (1983)

\bibitem{Visser1985}
M.~Visser, Phys. Lett. B \textbf{159}, 22 (1985)

\bibitem{Antoniadis1990}
I.~Antoniadis, Phys. Lett. B \textbf{246}, 377 (1990)

\bibitem{AntoniadisArkani-HamedDimopoulosDvali1998}
I.~Antoniadis, N.~Arkani-Hamed, S.~Dimopoulos, et~al., Phys. Lett. B
  \textbf{436}, 257 (1998), \eprint{hep-ph/9804398}

\bibitem{Arkani-HamedDimopoulosDvali1998a}
N.~Arkani-Hamed, S.~Dimopoulos, G.~Dvali, Phys. Lett. B \textbf{429}, 263
  (1998), \eprint{hep-ph/9803315}

\bibitem{RandallSundrum1999}
L.~Randall, R.~Sundrum, Phys. Rev. Lett. \textbf{83}, 3370 (1999),
  \eprint{hep-ph/9905221}

\bibitem{RandallSundrum1999a}
L.~Randall, R.~Sundrum, Phys. Rev. Lett. \textbf{83}, 4690 (1999),
  \eprint{hep-th/9906064}

\bibitem{AntoniadisArvanitakiDimopoulosGiveon2012}
I.~Antoniadis, A.~Arvanitaki, S.~Dimopoulos, et~al., Phys.Rev.Lett.
  \textbf{108}, 081602 (2012), \eprint{1102.4043}

\bibitem{YangLiuZhongDuWei2012}
K.~Yang, Y.-X. Liu, Y.~Zhong, et~al., Phys. Rev. D \textbf{86}, 127502 (2012),
  \eprint{1212.2735}

\bibitem{Rubakov2001}
V.~A. Rubakov, Phys. Usp. \textbf{44}, 871 (2001), \eprint{hep-ph/0104152}

\bibitem{Csaki2005}
C.~Csaki, in M.~Shifman, A.~Vainshtein, J.~Wheater, eds., \emph{From Fields to
  Strings: Circumnavigating Theoretical Physics : Ian Kogan Memorial
  Collection}, vol.~2, 967 (World Scientific, 2005), \eprint{hep-ph/0404096}

\bibitem{AppelquistChodosFreund1987}
T.~Appelquist, A.~Chodos, P.~G.~O. Freund, \emph{Modern Kaluza-Klein Theories}
  (Addison-Wesley Publishing Company, 1987)

\bibitem{DeWolfeFreedmanGubserKarch2000}
O.~DeWolfe, D.~Z. Freedman, S.~S. Gubser, et~al., Phys. Rev. D \textbf{62},
  046008 (2000), \eprint{hep-th/9909134}

\bibitem{Gremm2000}
M.~Gremm, Phys. Lett. B \textbf{478}, 434 (2000), \eprint{hep-th/9912060}

\bibitem{CsakiErlichHollowoodShirman2000}
C.~Csaki, J.~Erlich, T.~J. Hollowood, et~al., Nucl. Phys. B \textbf{581}, 309
  (2000), \eprint{hep-th/0001033}

\bibitem{AriasCardenasQuiros2002}
O.~Arias, R.~Cardenas, I.~Quiros, Nucl. Phys. B \textbf{643}, 187 (2002),
  \eprint{hep-th/0202130}

\bibitem{Barbosa-CendejasHerrera-Aguilar2005}
N.~Barbosa-Cendejas, A.~Herrera-Aguilar, J. High Energy Phys. \textbf{10}, 101
  (2005), \eprint{hep-th/0511050}

\bibitem{Barbosa-CendejasHerrera-Aguilar2006}
N.~Barbosa-Cendejas, A.~Herrera-Aguilar, Phys. Rev. D \textbf{73}, 084022
  (2006), \eprint{hep-th/0603184}

\bibitem{Barbosa-CendejasHerrera-AguilarReyesSchubert2008}
N.~Barbosa-Cendejas, A.~Herrera-Aguilar, M.~A. Reyes~Santos, et~al., Phys. Rev.
  D \textbf{77}, 126013 (2008), \eprint{0709.3552}

\bibitem{Buchdahl1970}
H.~A. Buchdahl, Mon. Not. Roy. Astron. Soc. \textbf{150}, 1 (1970)

\bibitem{BarrowOttewill1983}
J.~D. Barrow, A.~C. Ottewill, J. Phys. A \textbf{16}, 2757 (1983)

\bibitem{BarrowCotsakis1988}
J.~D. Barrow, S.~Cotsakis, Phys. Lett. B \textbf{214}, 515 (1988).

\bibitem{MukhanovKofmanPogosian1987}
V.~F. Mukhanov, L.~Kofman, D.~Y. Pogosian, Phys. Lett. B \textbf{193}, 427
  (1987)

\bibitem{CognolaElizaldeNojiriOdintsovSebastianiZerbini2008}
G.~Cognola, E.~Elizalde, S.~Nojiri, et~al., Phys. Rev. D \textbf{77}, 046009
  (2008).

\bibitem{DeTsujikawa2010}
A.~De~Felice, S.~Tsujikawa, Living Rev. Rel. \textbf{13}, 3 (2010),
  \eprint{1002.4928}

\bibitem{SotiriouFaraoni2010}
T.~P. Sotiriou, V.~Faraoni, Rev. Mod. Phys. \textbf{82}, 451 (2010),
  \eprint{0805.1726}

\bibitem{NojiriOdintsov2011}
S.~Nojiri, S.~D. Odintsov, Phys. Rept. \textbf{505}, 59 (2011),
  \eprint{1011.0544}

\bibitem{ParryPichlerDeeg2005}
M.~Parry, S.~Pichler, D.~Deeg, JCAP \textbf{0504}, 014 (2005),
  \eprint{hep-ph/0502048}

\bibitem{DeruelleSasakiSendouda2008}
N.~Deruelle, M.~Sasaki, Y.~Sendouda, Prog. Theor. Phys. \textbf{119}, 237
  (2008), \eprint{0711.1150}

\bibitem{BalcerzakDabrowski2009}
A.~Balcerzak, M.~P. Dabrowski, JCAP \textbf{0901}, 018 (2009),
  \eprint{0804.0855}

\bibitem{BalcerzakDabrowski2008}
A.~Balcerzak, M.~P. Dabrowski, Phys. Rev. D \textbf{77}, 023524 (2008),
  \eprint{0710.3670}

\bibitem{HoffDias2011}
J.~Hoff~da Silva, M.~Dias, Phys. Rev. D \textbf{84}, 066011 (2011),
  \eprint{1107.2017}

\bibitem{CaramesGuimaraesHoff2013}
T.~Carames, M.~Guimaraes, J.~Hoff~da Silva, Phys. Rev. D \textbf{87}, 10,
  106011 (2013), \eprint{1205.4980}

\bibitem{AfonsoBazeiaMenezesPetrov2007}
V.~I. Afonso, D.~Bazeia, R.~Menezes, et~al., Phys. Lett. B \textbf{658}, 71
  (2007), \eprint{0710.3790}

\bibitem{LiuZhongZhaoLi2011}
Y.-X. Liu, Y.~Zhong, Z.-H. Zhao, et~al., J. High Energy Phys. \textbf{06}, 135
  (2011), \eprint{1104.3188}

\bibitem{BazeiaMenezesPetrovSilva2013}
D.~Bazeia, R.~Menezes, A.~Y. Petrov, et~al., Phys. Lett. B \textbf{726}, 523
  (2013), \eprint{1306.1847}

\bibitem{BazeiaLobaoMenezesPetrovSilva2014}
D.~Bazeia, A.~S. Lob\~ao, R.~Menezes, et~al., Phys. Lett. B \textbf{729}, 127
  (2014), \eprint{1311.6294}

\bibitem{XuZhongYuLiu2015}
Z.-G. Xu, Y.~Zhong, H.~Yu, et~al., Eur. Phys. J. C \textbf{75}, 8, 368 (2015),
  \eprint{1405.6277}

\bibitem{GuGuoYuLiu2014}
B.-M. Gu, B.~Guo, H.~Yu, et~al., Phys. Rev. D \textbf{92}, 024011 (2015),
  \eprint{1411.3241}

\bibitem{BazeiaLobaoMenezes2015}
D.~Bazeia, A.~Lob\~ao, R.~Menezes, Phys. Lett. B \textbf{743}, 98 (2015),
  \eprint{1502.04757}

\bibitem{BazeiaLobaoLosanoMenezesOlmo2015}
D.~Bazeia, A.~Lob\~ao, L.~Losano, et~al., Phys. Rev. D \textbf{91}, 12, 124006
  (2015), \eprint{1505.06315}

\bibitem{YuZhongGuLiu2015}
H.~Yu, Y.~Zhong, B.-M. Gu, et~al.  (2015), \eprint{1506.06458}

\bibitem{ZhongLiuYang2011}
Y.~Zhong, Y.-X. Liu, K.~Yang, Phys. Lett. B \textbf{699}, 398 (2011),
  \eprint{1010.3478}

\bibitem{DzhunushalievFolomeevKleihausKunz2010}
V.~Dzhunushaliev, V.~Folomeev, B.~Kleihaus, et~al., J. High Energy Phys.
  \textbf{04}, 130 (2010), \eprint{0912.2812}

\bibitem{LiuLuWang2012}
H.~Liu, H.~Lu, Z.-L. Wang, J. High Energy Phys. \textbf{1202}, 083 (2012),
  \eprint{1111.6602}

\bibitem{Giovannini2001a}
M.~Giovannini, Phys. Rev. D \textbf{64}, 064023 (2001), \eprint{hep-th/0106041}

\bibitem{Giovannini2002}
M.~Giovannini, Phys. Rev. D \textbf{65}, 064008 (2002), \eprint{hep-th/0106131}

\bibitem{Giovannini2003}
M.~Giovannini, Classical Quantum Gravity \textbf{20}, 1063 (2003),
  \eprint{gr-qc/0207116}

\bibitem{Carroll2004}
S.~Carroll, \emph{Spacetime And Geometry An Introduction To General Relativity}
  (Pearson Education, Inc., 2004)

\bibitem{ZhongLiu2013}
Y.~Zhong, Y.-X. Liu, Phys. Rev. D \textbf{88}, 024017 (2013),
  \eprint{1212.1871}

\bibitem{FierzPauli1939}
M.~Fierz, W.~Pauli, Proc. Roy. Soc. Lond. \textbf{A173}, 211 (1939)

\bibitem{Bardeen1980}
J.~M. Bardeen, Phys. Rev. D \textbf{22}, 22 (1980)

\bibitem{KodamaSasaki1984}
H.~Kodama, M.~Sasaki, Progr. Theoret. Phys. Suppl. \textbf{78}, 1 (1984)

\bibitem{MukhanovFeldmanBrandenberger1992}
V.~F. Mukhanov, H.~A. Feldman, R.~H. Brandenberger, Phys. Rep. \textbf{215},
  5-6, 203 (1992).

\bibitem{Weinberg2008}
S.~Weinberg, \emph{Cosmology} (Oxford University Press, 2008)

\bibitem{BazeiaGomesLosano2009}
D.~Bazeia, A.~R. Gomes, L.~Losano, Int. J. Mod. Phys. A \textbf{24}, 1135
  (2009), \eprint{0708.3530}

\end{thebibliography}

\end{document}